\documentclass[journal]{IEEEtran}
\usepackage{cite}
\usepackage{amsmath,amssymb,amsfonts}
\usepackage{algorithmic}
\usepackage{graphicx}

\usepackage{caption}
\usepackage{subcaption}

\ifCLASSINFOpdf
\else
\fi
\begin{document}
%
\title{Development of a Predictive Process Design kit for 15-nm FinFETs: FreePDK15}
%
%
%

\author{Kirti Bhanushali, Chinmay Tembe, and W. Rhett Davis  
}

\maketitle

\begin{abstract}
FinFETs are predicted to advance semiconductor scaling for sub-20nm devices. In order to support their introduction into research and universities it is crucial to develop an open source predictive process design kit. This paper discusses in detail the design process for such a kit for 15nm FinFET devices, called the FreePDK15. The kit consists of a layer stack with thirteen-metal layers based on hierarchical-scaling used in ASIC architecture, Middle-of-Line local interconnect layers and a set of Front-End-of-Line layers. The physical and geometrical properties of these layers are defined and these properties determine the density and parasitics of the design. The design rules are laid down considering additional guidelines for process variability, challenges involved in FinFET fabrication and a unique set of design rules are developed for critical dimensions. Layout extraction including modified rules for determining the geometrical characteristics of FinFET layouts are implemented and discussed to obtain successful Layout Versus Schematic checks for a set of layouts. Moreover, additional parasitic components of a standard FinFET device are analyzed and the parasitic extraction of sample layouts is performed. These extraction results are then compared and assessed against the validation models.
\end{abstract}

\begin{IEEEkeywords}
FreePDK, FinFET 15nm, Process Design Kit, Middle-of-Line layers, DRC, LVS, parasitic extraction.
\end{IEEEkeywords}

%
\IEEEpeerreviewmaketitle

\section{Introduction}
The International Technology Roadmap for Semiconductors (ITRS) forecasts the physical length of the transistors to scale down to 16nm by 2016 \cite{ITRS:Online}. However, the scaling of bulk MOS technology for sub-20nm transistors has faced major problems - these include high leakage power, random dopant fluctuations, Drain-Induced-Barrier-Lowering (DIBL) and other short channel effects. An alternative device called a FinFET has emerged, and it has been demonstrated to advance scaling of seminconductor technology beyond 20-nm processes. FinFETs achieve lower sub-threshold leakage and improved short channel characteristics due to an advanced three dimensional multi-gate geometry. Further, due to an improved gate control and a depleted thin fin structure, they achieve better short channel performance and have lower random dopant fluctuation \cite{RDF}.

The Process design kits (PDKs) for these technologies have already been developed for commercial FinFET processes at the 14 nm. However, these processes are not readily available for the university education purposes due to the critical nature of intellectual property. Additionally, a large investment is required for licensing these processes, which is beyond the scope of universities. Thus, there is an immediate need for development of an open source predictive process design kit to help students gain a detailed understanding of standard design processes. The development of FreePDK15, is thus a step towards achieving a complete predictive process design. 
 
In this paper, the development of an entire metal layer stack based on the ITRS \cite{ITRS:Online} estimates has been presented. The standard FinFET layout is evaluated from the point of view of fabrication and design rules. Additionally, due to limitations of the standard photo-lithography processes, state-of-the-art techniques like double-patterning lithography have been assumed for critical dimensions. Moreover, cut masks and Middle-of-Line (MOL) layers are facilitated to enhance cell density and thus require special design rules. To validate these design rules, layouts of an Inverter, NAND4, and their cascaded cells have been designed and their density is evaluated. A set of the formulae required to accurately identify and calculate the source and drain dimensions of the FinFET layout are presented. Additionally, layout extraction rules for double patterning, metal stitching and gate cut layers are implemented and validated. The parasitic characteristics of a standard FinFET device are studied and the extraction rules for parasitic capacitance and resistance are implemented. The extraction of a set of standard layouts is compared and assessed against first order capacitance and resistance models for metal layers.

This paper discusses the intermediate steps involved in the development of process design kit FreePDK15. In section II, the layers used for the PDK are discussed. In section III, a standard FinFET layout cell is presented and the design rules for these layouts are explained in section IV. Section V discusses steps involved in layout extraction. In section VI parasitic extraction and validation is discussed and the paper is concluded in section VII.

\section{FreePDK15 Layer stack}
The layer stack for FreePDK15 is developed considering multiple factors, including multi-pattern lithography, metal stitching, dense routing and improvement of contact resistances. The layer stack includes additional layers to accommodate for the three-dimensional nature of the FinFET device and the layer properties follow the predictions from the International Technology Roadmap for Semiconductors 2011 for the 2016 node \cite{ITRS_interconnect}.

\begin{figure}[htbp]
\centering
\includegraphics[width=0.48\textwidth]{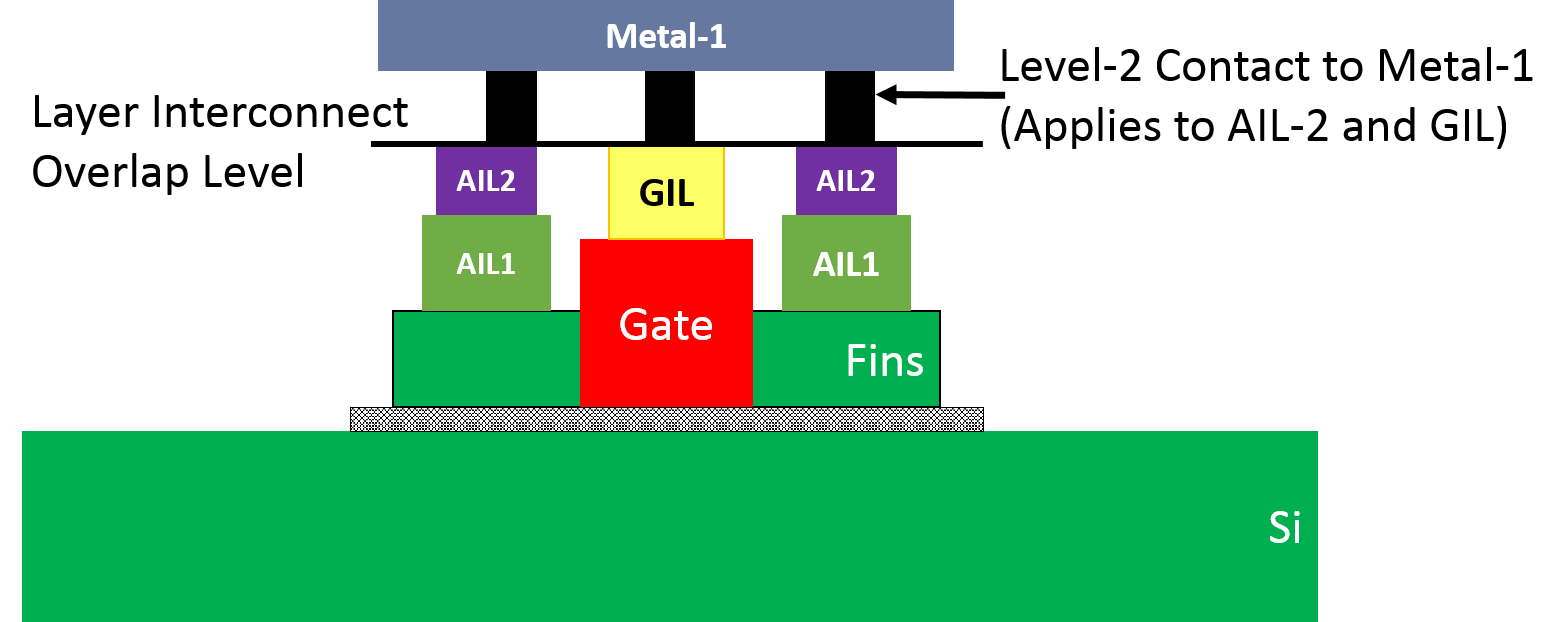}
\DeclareGraphicsExtensions{.pdf,.jpeg,.png,.jpg} 
\caption{Cross Section of a FinFET device}
\label{figure_cross-section}
\end{figure} 
The standard cross-section for a FinFET can be seen in Figure \ref {figure_cross-section}. The cross-section indicates the use of Middle-Of-Line (MOL) layers along with the standard Back-End-Of-Line (BEOL) layers, and Front-End-Of-Line(FEOL) layers. 

\subsection{BEOL Layers}
The ITRS-2011 tables for the 2016 node predicts use of 13 metal layers. The metal layer stack thus includes 13 layers and follows hierarchical scaling used for standard ASIC architecture \cite{ITRS_interconnect}. This is further divided into Metal1 layer, Intermediate metal layers, Semi-Global metal layers and Global metal layers.

\begin{figure}[htbp]
\centering
\includegraphics[width = 0.48\textwidth]{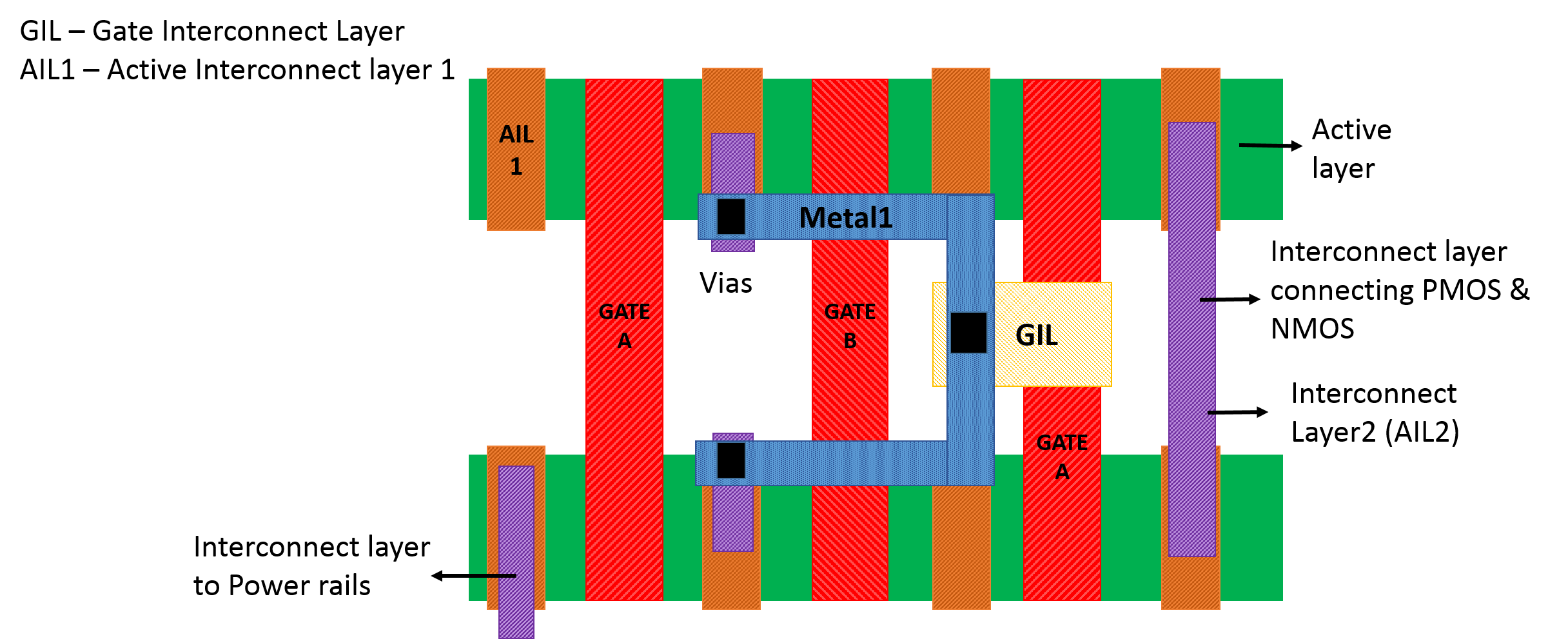}
\DeclareGraphicsExtensions{.pdf,.jpeg,.png,.jpg} 
\caption{ MOL layers used as interconnects  }
\label{figure_interconnections}
\end{figure}

\begin{table*}[tbh]
\centering
\caption{MOL layers and their functions}
\begin{tabular}{|p{5cm}|p{6cm}|}\hline
Layer&Purpose\\ \hline
Active Interconnect Layer-1 (AIL1) & Connecting individual fins\\ \hline
Active Interconnect Layer-2 (AIL2) & Connecting AIL-1 to Metal-1 through a via\\ \hline
Gate Interconnect Layer (GIL) & Connecting Gate to Metal1\\ \hline
\hline\end{tabular}
\label{MOL_layers}
\end{table*}

\subsection{Middle-of-Line Layers}
\label{mol}
The MOL layers act as an interface between FEOL layers like ACT and BEOL layers like Metal1. The MOL layers are implemented for overcoming electrical resistance concerns and the loss of performance between inter-connected layers \cite{design_rules}. Although the concept of MOL layers has been studied in the past the  primary inspiration behind their use for FreePDK15 comes from \cite{Schuddinck:FinFET}. In  \cite{Schuddinck:FinFET} 14nm bulk FinFET standard cells have been implemented and the impact of MOL layers, local interconnect layers IM1 and IM2, on cell parasitics is analyzed. MOL layers can be used for connecting internal nets as indicated in Figure \ref{figure_interconnections}. Thus these layers help in achieving denser layouts with the provision for connecting internal nets, internal devices as well as providing connection to the power rails. This eliminates the use of Metal1 layer for internal routing and thus that of additional contacts/vias.The function of all MOL layers is listed in Table \ref{MOL_layers}.

\subsection{Double patterning and other techniques}
\subsubsection{Double patterning}
Fabrication beyond 20nm involves multiple challenges, primarily from the standpoint of photo-lithography as the gate pitch is much smaller. It is very difficult to fabricate devices using the standard 193nm Argon Flouride (ArF) lasers. We also assume that Extreme Ultra-Violet (EUV) lithography is not used in this process because it has not been able to achieve the desired yield for volume production. Instead, double patterning is the technique assumed for FreePDK15, since it achieves greater pitch density compared to standard single-patterning lithography. This is due to superior contrast obtained from exposed and unexposed areas. In principle, the layout is decomposed into two masks, each with different colors each with half the pattern to be printed. In FreePDK15 this is  implemented by providing two different colored layers for layers with critical dimensions, like lower metal layers and gate layer.

For example, in FreePDK15, double patterning is assumed for gate layers. GATEA and GATEB are two differently colored gate layers

\subsubsection{Cut layers}
In addition to these layers, FreePDK15 also consists of a Gate Cut mask/cut layer called GATEC to remove unwanted features printed by its preceding  mask. This helps in printing non-uniform device structures and in overcoming errors due to mask misalignment. It is used to break connectivity between gate layers that are continuous. It is very convenient to form long GATEA/GATEB/GATEAB shapes and then create multiple individual gate shapes by using a grid of GATEC layers, rather than patterning multiple gate shapes at specific places on the wafer.

\section{FinFET Layout abstractions/approaches}
\label{sec:fin_layout}
As the FinFET device has three-dimensional thin fin sturcture, it requires additional fabrication steps compared to a standard planar MOSFET. These differences are primarily due to width quantization and use of MOL Layers.

\begin{figure}[!ht]
\centering
\includegraphics[width = 0.48\textwidth]{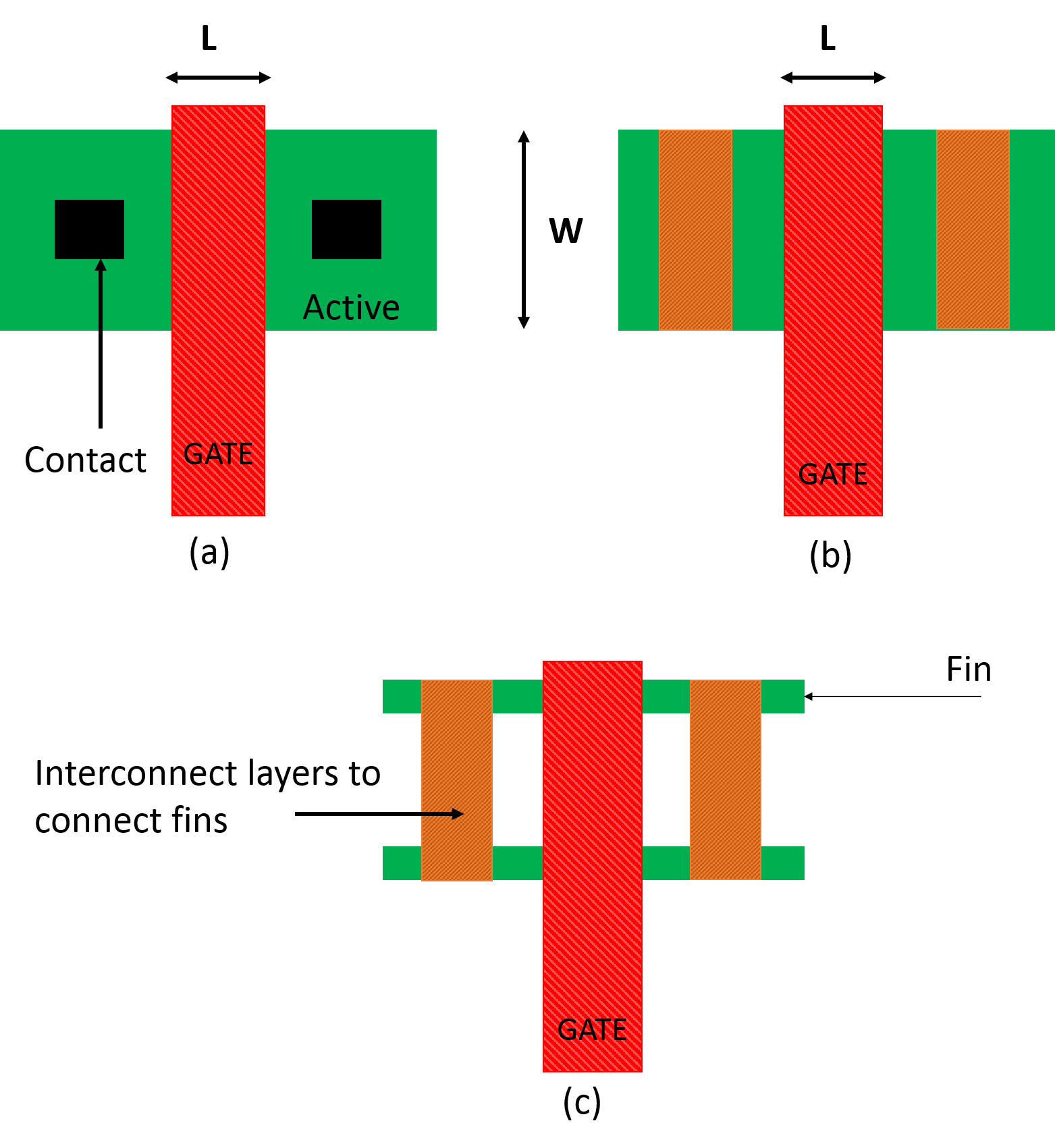}
\DeclareGraphicsExtensions{.pdf,.jpeg,.png,.jpg} 
\caption{Basic transistor layouts (a) Planar MOS (b) FinFET (c) Physical Mask - FinFET }
\label{fig_single_cell}
\end{figure}
\subsection{Single FinFET transistor} 
The layout of a single planar MOS transistor with width W and gate length L is presented in Fig \ref{fig_single_cell}.a. The active layer has a direct contact to Metal1 layer for the planar MOS. But, in the layout of a FinFET transistor, contact is established through local interconnect layers. Figure \ref{fig_single_cell}.b shows the representation of the FinFET layout drawn in the design tool, however,  due to the quantization of the fin width the device structure on the physical mask looks different and is illustrated in Fig.\ref{fig_single_cell}
\begin{figure}
\centering
\includegraphics[width=0.30\textwidth]{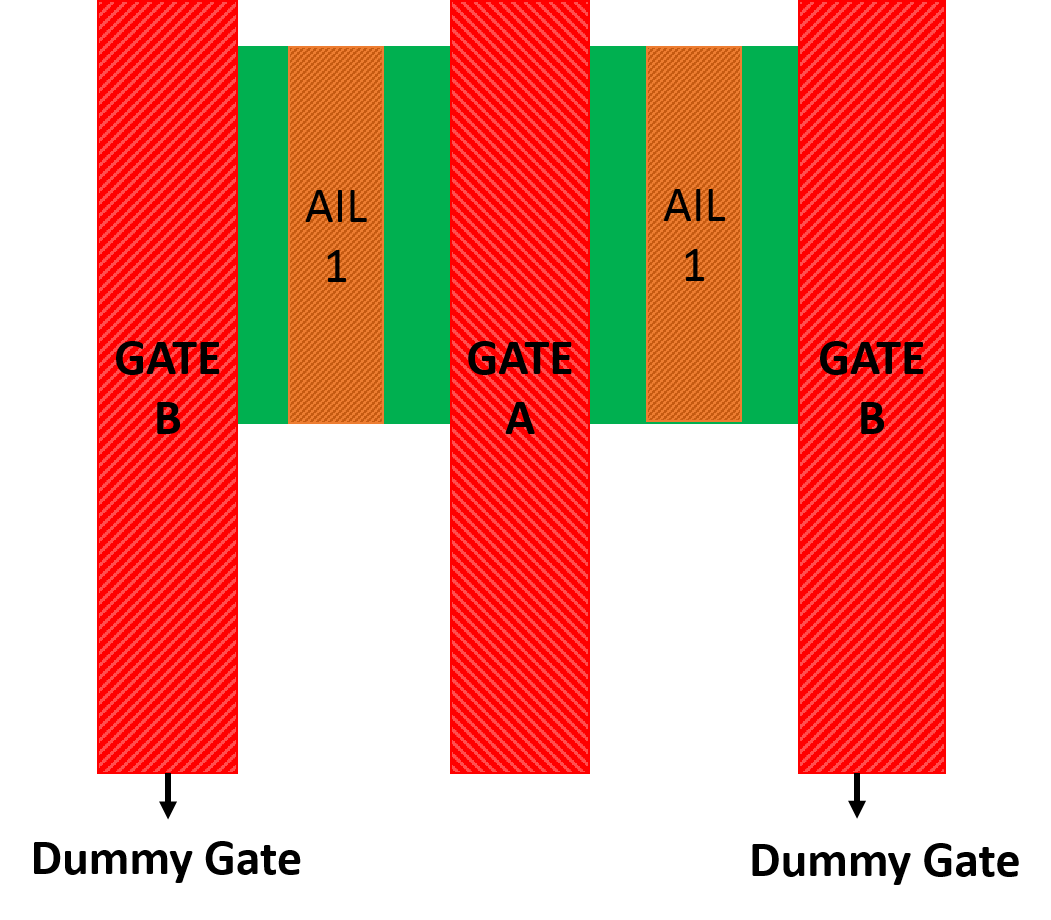}
\DeclareGraphicsExtensions{.pdf,.jpeg,.png,.jpg} 
\caption{Dummy gates for process uniformity and DPL}
\label{fig_dummy}
\end{figure}

In order to ensure process uniformity \cite{Tom_dillinger} in sub-20nm transistors ``dummy'' gates are also printed at the end of the Fins as seen in Figure \ref{fig_dummy}. Moreover, GATEA and GATEB have different patterns for double patterning lithography

\section{Design rule development}
The design rules define the basic geometric and connectivity restrictions for a device technology and are thus important to its development. They ensure sufficient margins against manufacturing process variability. In addition to that they help the designer in verification of the design before it is sent for fabrication. Violations of these rules can result in undesirable operation of the circuits, thus they are critical to the circuit reliability. Furthermore, these design rules are crucial in defining the density of the integrated circuit as non-optimum design rules would result in wastage of critical design space. Also, with the use of emerging technologies like FinFETs, it is necessary to introduce new sets of design rules to efficiently achieve correct functionality.
 
Typically, the number of design rules can vary from few hundreds to thousands. The design rules for FreePDK15 are implemented considering the geometric, electrical and lithographic constraints. They incorporate standard minimum width and spacing rules, along with certain restrictive design rules. 

\subsection{Standard design rules}
The standard design rules are listed in Table II
\begin{table}[!h]
\centering
\label{table_des}
\caption{Standard design rules and their functions}
\begin{tabular}{|p{2cm}|p{5.5cm}|}\hline
Rule&Function\\ \hline
Minimum width & Defined by the resolution of the lithographic process used, prevents open-circuits.\\ \hline
Minimum Spacing & Ensure electrical isolation between two shapes\\ \hline
Enclosure & Prevent overlay errors due to misalignment of layers\\ \hline
Overlap & Ensure reliability during misalignment of layers\\ \hline
Area & Ensure adhesion and prevent overlay errors\\ \hline  
\hline\end{tabular}
\end{table}

\subsection{Advanced design rules}
These rules are specific to FinFET layout and double patterning lithography. 

\begin{figure}[!h]
\centering
\includegraphics[width=0.48\textwidth]{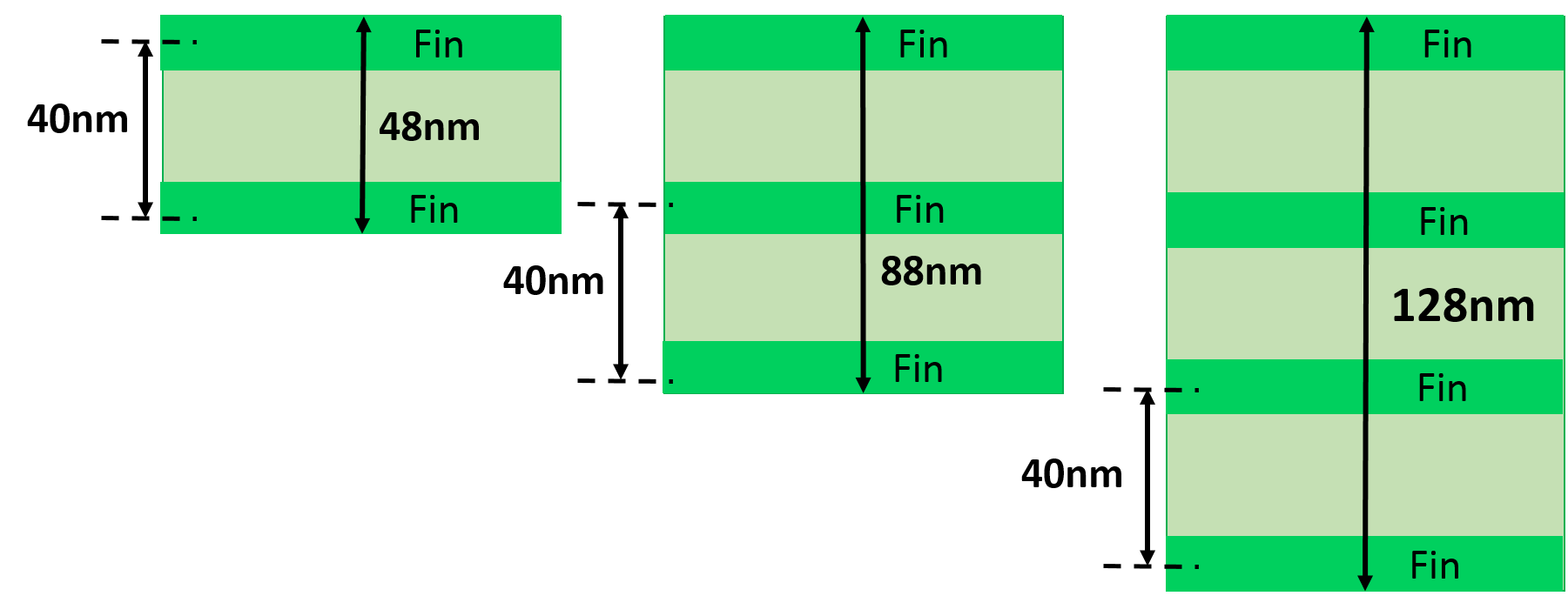}
\DeclareGraphicsExtensions{.pdf,.jpeg,.png,.jpg} 
\caption{Width quantization of the active layer}
\label{figure_width_quantization}
\end{figure} 
\subsubsection{Incremental width rule: Active}
The incremental width rule is introduced due to the discrete nature of the FinFET width. The total width of a FinFET device is defined by the number of fins in the device and thus can only increase in discrete steps. As it can be seen in  Figure \ref{figure_width_quantization} the active width can only increment in steps of 40nm, which is the pitch of the active layer  \cite{Bhanushali:2015}.

\begin{figure}[!h]
\centering
\includegraphics[width=0.48\textwidth]{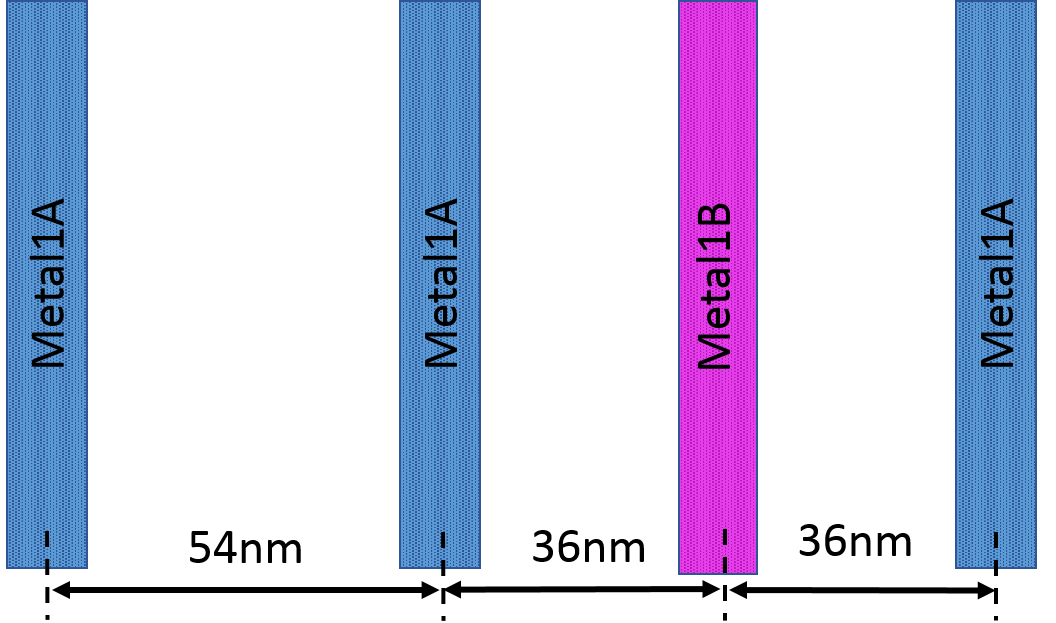}
\DeclareGraphicsExtensions{.pdf,.jpeg,.png,.jpg} 
\caption{Different pitch rules for different layers}
\label{figure_double_patterning}
\end{figure} 

\subsubsection{Multi-colored design rules}
A distinct feature of device fabrication in the sub-20nm technology is implementation of multi-patterning lithography. In FreePDK15 double patterning lithography (DPL) is assumed. It is thus necessary to use different rules for different colored metal layers. Figure \ref{figure_double_patterning} shows that the required minimum pitch between two similar metal layers, Metal1A layers in this case, is bigger than the minimum required pitch between metal layers of different colors, Metal1A and Metal1B.

\subsubsection{Restrictive design rules}
Restrictive design rules are introduced to maintain the conventional design methodologies with introduction of a new set of restrictions. An example of restrictive design rule is allowing only discrete gate lengths. Another example is restriction of jogs and bends in gate layers as it can result in pinching \cite{Davis_book}. However, as this rule causes an increase in the overall area of the layout, it is only implemented for critical dimensions.

\subsection{Design rule validation} 
The design rules for FreePDK15 are predictive at best and need further validation. A set of layouts were drawn and a design rule check was performed on them for validating these rules \cite{FreePDK, } .

\begin{figure}
\centering
\includegraphics[width=0.48\textwidth]{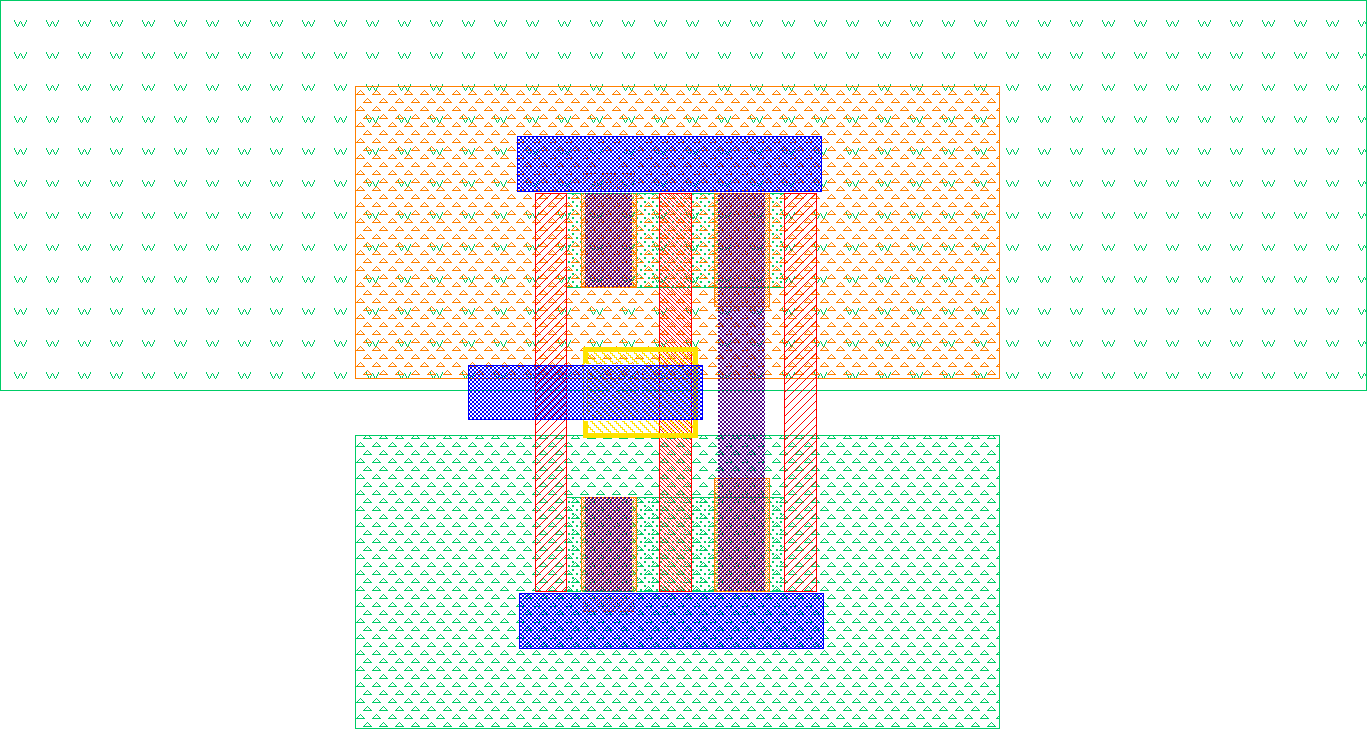}
\DeclareGraphicsExtensions{.pdf,.jpeg,.png,.jpg} 
\caption{Standard inverter cell-FreePDK15}
\label{fig_Inverter}
\end{figure}

\subsubsection{Inverter cell}
A standard minimum sized FreePDK15 Inverter cell is presented in Figure \ref{fig_Inverter}. It uses AIL-2 for connecting the internal nets and power rails. Additionally, GATEA and GATEB are implemented for process uniformity \cite{FreePDK, Thesis}. 

\begin{figure}[!ht]
\centering
\includegraphics[width=0.48\textwidth]{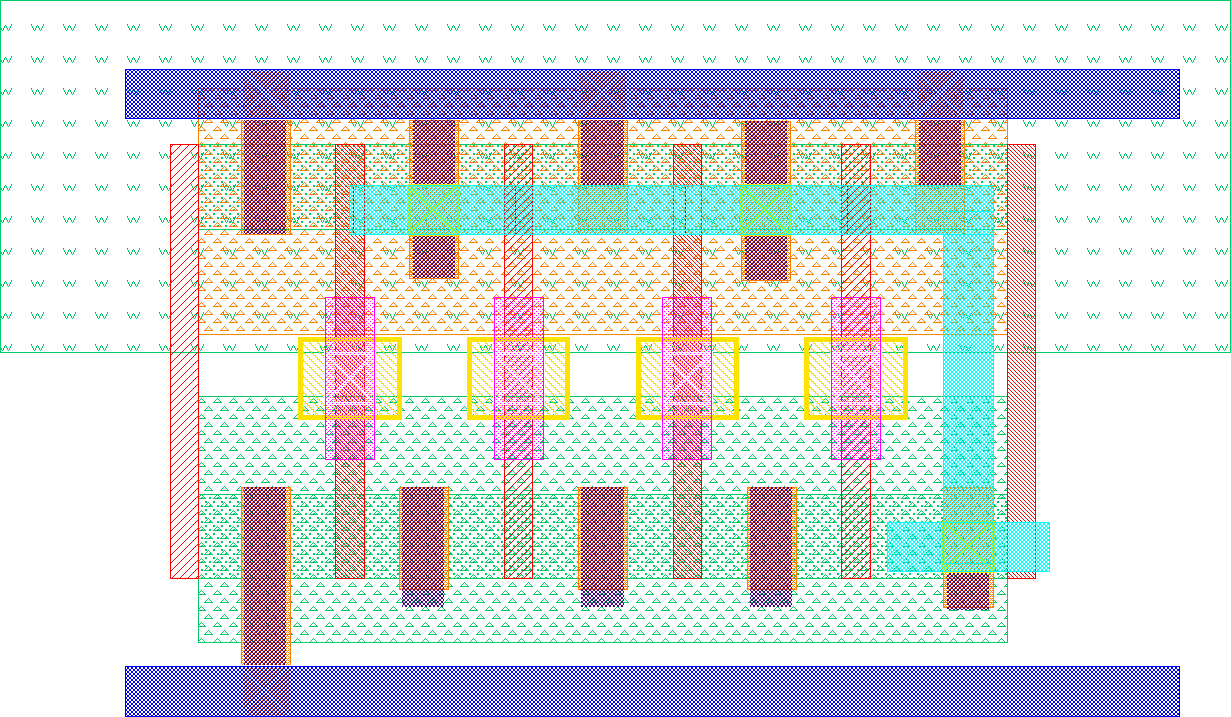}
\DeclareGraphicsExtensions{.pdf,.jpeg,.png,.jpg} 
\caption{Standard NAND4 cell-FreePDK15}
\label{fig_NAND4}
\end{figure}

\subsubsection{NAND4 cell}
A standard NAND4 cell shown in Figure \ref{fig_NAND4} consists of double colored metal1 layer for layout density; the design rules were further validated by running design rule checks. 

\begin{figure}[!h]
\centering
\includegraphics[width=0.48\textwidth]{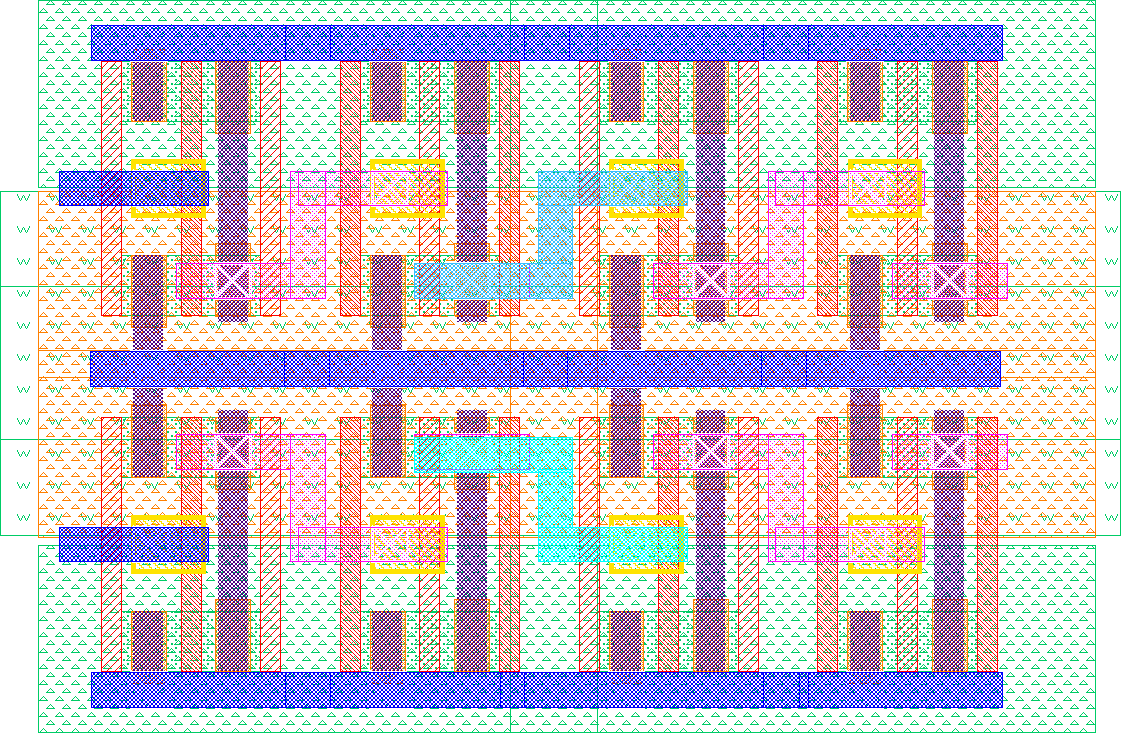}
\DeclareGraphicsExtensions{.pdf,.jpeg,.png,.jpg} 
\caption{Tiled Inverter layout-FreePDK15}
\label{fig_tiled_inverter}
\end{figure}

\begin{figure}[!h]
\centering
\includegraphics[width=0.40\textwidth]{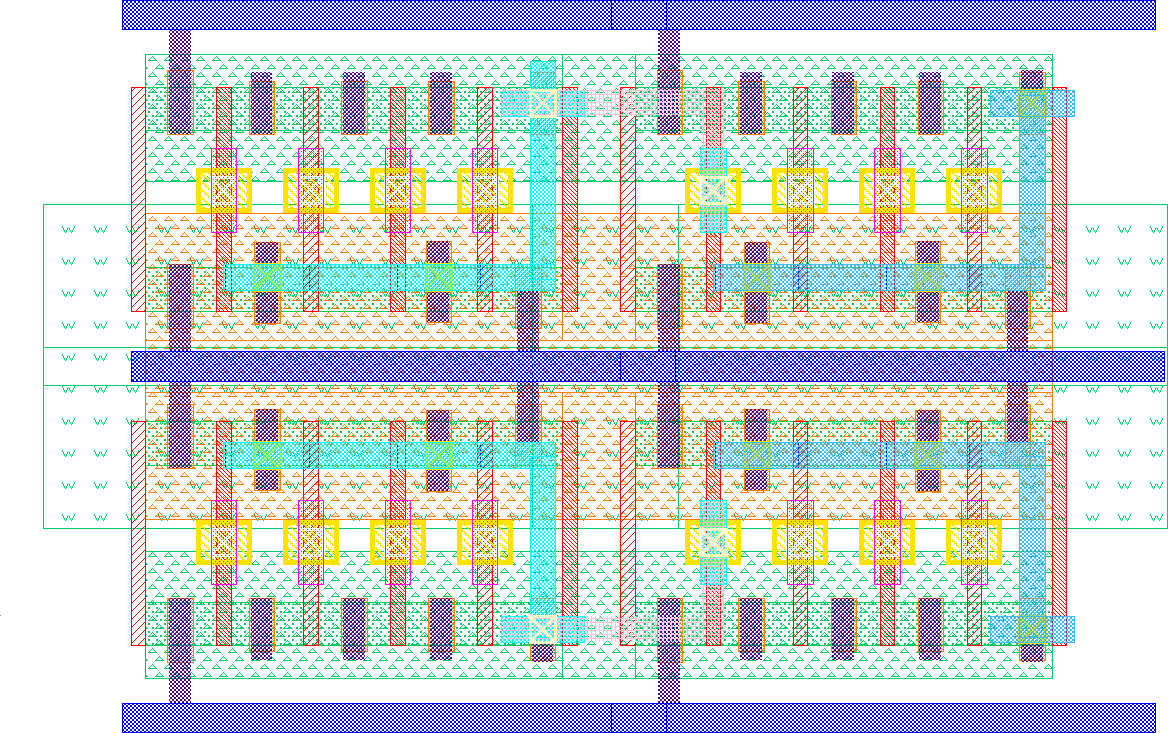}
\DeclareGraphicsExtensions{.pdf,.jpeg,.png,.jpg} 
\caption{Tiled NAND4 layout-FreePDK15}
\label{fig_tiled_NAND}
\end{figure}

\begin{figure}[!h]
\centering
\includegraphics[width=0.40\textwidth]{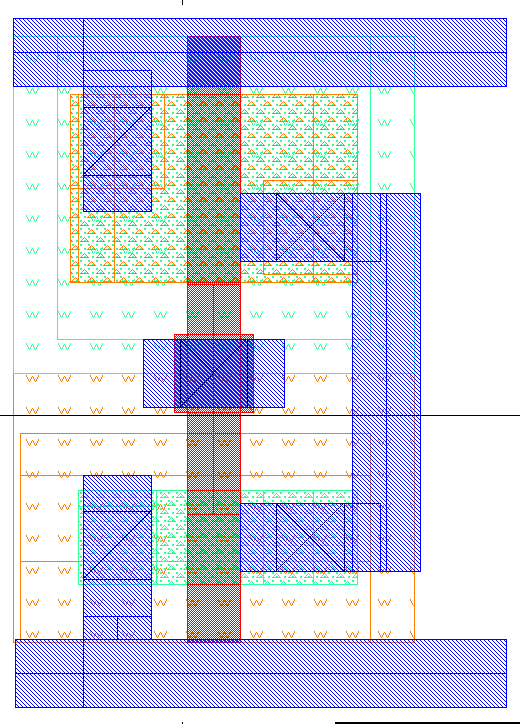}
\DeclareGraphicsExtensions{.pdf,.jpeg,.png,.jpg} 
\caption{Minimum sized inverter for 45nm bulk CMOS technology(FreePDK45)}
\label{fig_Inverter_45}
\end{figure}

\subsubsection{Tiled cells}
Tiled Inverter and NAND4 layouts presented in Figure \ref{fig_tiled_inverter} and Figure \ref{fig_tiled_NAND} resp. are also designed for validating the design rules of higher order metal layers.

\subsubsection{Layout density comparison}
The area of minimum sized FinFET inverter is compared with the standard bulk MOS inverter designed using 45nm bulk FreePDK45 \cite{FreePDK45} \cite{FreePDK} process in order to evaluate the layout density of the FinFET process.  The layout density in FinFETs does not scale as in bulk MOSFETs.  In \cite{Alioto} the layout density for a FinFET design is found to be 1.3 times that for the bulk process at the same process node of 65nm. The primary reason for this can be attributed to the area overhead and width quantization issue in FinFETs. 

\begin{figure*}[htbp]
\centering
\begin{minipage}{.5\textwidth}
  \centering
  \includegraphics[width=.9\linewidth]{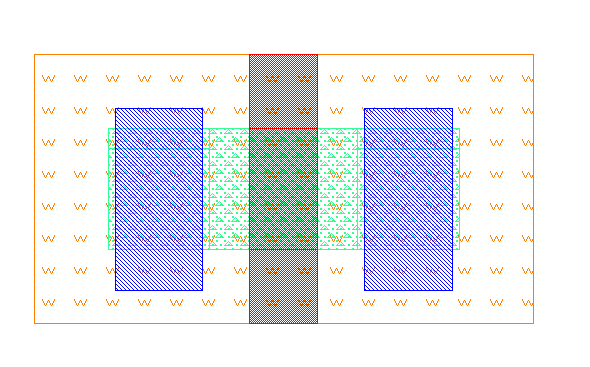}
  \subcaption{NMOS Layout: FreePDK45}
  \label{fig:sub1}
\end{minipage}%
\begin{minipage}{.5\textwidth}
  \centering
  \includegraphics[width=.8\linewidth]{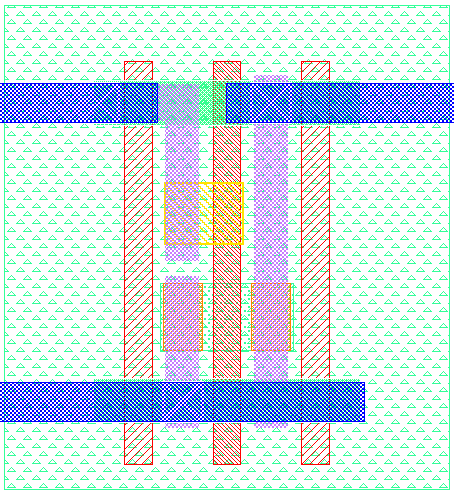}
  \subcaption{ NFinFET Layout: FreePDK15}
  \label{fig:sub2}
\end{minipage}
\caption{Comparison of NFinFET and NMOS layout}
\label{fig:both}
\end{figure*}

The area of a standard CMOS bulk technology(FreePDK45) as shown in Figure \ref{fig_Inverter_45} was compared with the FeePDK Inverter. The area shrink factor of 45nm CMOS inverter to 15nm FinFET inverter was found to be around 1/6.

\section{Layout extraction and Device recognition}
\label{sec:layout_ext}
A crucial element in the development of the process design kit is error-free layout extraction. Layout extraction involves both device recognition and connectivity extraction, and its output is a netlist that contains connectivity information of all the recognized devices. Thus, layout extraction rules for FreePDK15 have been developed for transistor devices NFinFETs and PFinFETs, however, no passive structures are defined yet \cite{Chinmay}.

\subsection{Device recognition: Bulk MOS vs FinFET} 
The shift from traditional bulk planar CMOS devices to FinFETs cause problems in device recognition. In contrast to the planar devices, FinFETs have a three dimensional folding of gate layer over the fin which adds to the complexity of creating layouts. However, as indicated in section \ref{sec:fin_layout}, the layout of a FinFET device is drawn similar to planar devices with a few exceptions. The comparison of a standard NMOS layout in FreePDK45 and an NFinFET in FreePDK15 is shown in Figure \ref{fig:both}. However, this doesn't account for the multi-fin nature of the FinFETs which results in modification of the formulae used for calculating source and drain dimensions of the FinFET device. Additionally, the gate length is only restricted to 14, 16 and 20 nm. In most cases a single length of 16nm would be enforced, but it is possible that the critical dimensions of all devices may be lengthened to 20 nm or shortened to 14 nm across the entire wafer.

%

It is also very important to correctly extract the drain and source dimensions i.e. the area, and perimeter, as they define the parasitic source and drain capacitances. 

\subsubsection{Source and Drain dimensions: Planar MOS}
The formulae used for estimating the areas (\textit{A\textsubscript{D}}, \textit{A\textsubscript{S}}) and the perimeters (\textit{P\textsubscript{D}}, \textit{P\textsubscript{S}}) of source and drain for planar devices from \cite{Rabaey} are given below.

\begin{equation}
\label{eqn:01}
A_D = A_S = W*L_{D/S}
\end{equation}

\begin{equation}
\label{eqn:02}
P_D = P_S = 2*L_{D/S}+ W
\end{equation}

\subsubsection{Source and Drain dimensions: FinFET}
For bulk FinFET devices, the drain and source area is represented by \textit{ADEJ}, and \textit{ASEJ} respectively, while the perimeter of the drain and source is represented by \textit{PDEJ}, and  \textit{PSEJ}. The formulae for these parameters account for the number of fins and are as shown in following equations \cite{BSIM_manual}.

\begin{equation}
\label{eqn:03}
ADEJ = ASEJ = n_{fin}*W_{fin}*L_{fin(D/S)}
\end{equation}

\begin{equation}
\label{eqn:04}
PDEJ = PSEJ = 2*L_{fin(D/S)}*n_{fin} + W_{fin}*n_{fin}
\end{equation}

\subsection{Layout Extraction rules}
In order to accurately extract a layout, a layout vs schematic (LVS) rule file is defined. Its accuracy depends on the accuracy of the rule file with regards to device definition and extraction, and connectivity extraction. These rules are validated by creating sample layouts and performing LVS checks on sample layouts.

FreePDK15 primarily follows the same LVS rules as defined for bulk MOS technology, however, due to the definition of MOL layers and introduction of cut-layers some of these rules are modified.

\subsection{MOL connectivity rules}
Due to the introduction of the MOL layers the device contact rules are modified. As indicated in section \ref{mol} AIL1 acts as the first local interconnect to active while GIL acts as the first local interconnect to Gate. However, AIL2 can act as local interconnect layer to AIL1 as well as GIL. This provides multiple device contact options: AIL1 - AIL2 - M1, GIL -M1 or GIL - AIL2 - M1.

\begin{figure}[!h]
\centering
\includegraphics[width = 0.48\textwidth]{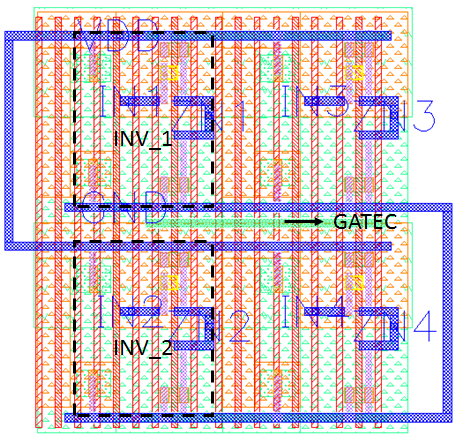}
\DeclareGraphicsExtensions{.pdf,.jpeg,.png,.jpg} 
\caption{ GATEC for breaking connectivity between Inverters}
\label{fig_gatec}
\end{figure}  

\subsection{Gate Cut rules}
Gate cut layer act as a negative mask and additional rules are defined to identify break in connectivity if a gate cut (GATEC) layer is present. As shown in Figure \ref{fig_gatec}, use of GATEC facilitates denser layouts in case of four tiled inverters by breaking GATE connectivity where required.

\subsection{Double patterning rules}
BEOL rules concern the way in which metal layers are connected. The connection of various metal layers is through the alternating via layers, as can be seen in the metal layer stack. For multiple patterned layers, all layers at the same level in the hierarchy, even with different colors, are treated as identical for layout extraction and can be connected to any of the multiple patterned layers of higher or lower levels of the hierarchy using the corresponding via. For example, intermediate metal layer MINT3, MINT3A, MINT3B are considered same and either of them can be connected to either of MINT4, MINT4A or MINT4B using via VINT3. Similarly, they can be connected to either of MINT2, MINT2A or MINT2B using via VINT2.

\begin{figure*}
\centering
\begin{minipage}{.45\textwidth}
  \includegraphics[width=.90\textwidth]{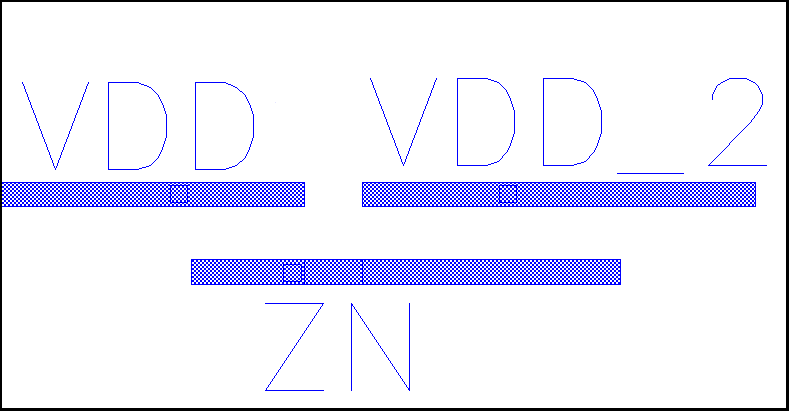}
  \centering
  \subcaption{Spacing violations between VDD and VDD$\_$2 preventing denser layouts}
  \label{fig:stitch1}
\end{minipage}%
\begin{minipage}{.45\textwidth}
  \includegraphics[width=.85\textwidth]{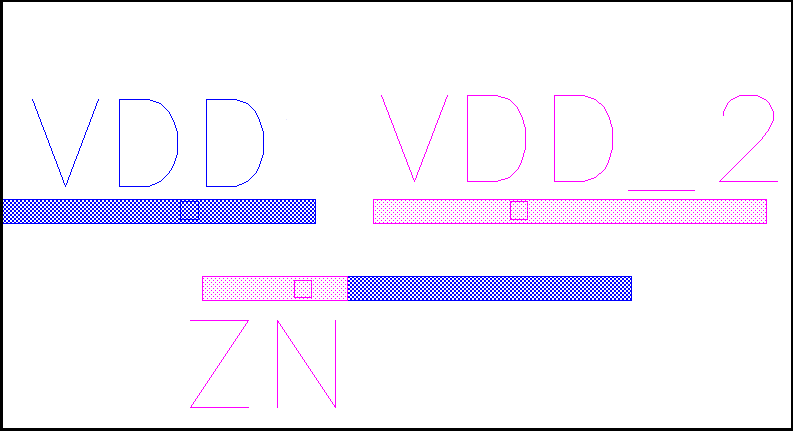}
  \centering
  \subcaption{Metal stitching permitting denser layouts due to use of different color of VDD, ZN and VDD$\_$2 nets}
  \label{fig:stitch2}
\end{minipage}
\caption{ Metal stitching resolves spacing violations and facilitates denser layouts}
\label{fig:both_stitch}
\end{figure*}

FreePDK15 has been developed to allow metal stitching, which is a means to connect multiple patterned layers to each other in order to save area. For example, MINT5 can connect to both MINT5A and MINT5B and vice versa. In situations where there are multiple violations to design rules, specifically spacing rules, instead of modifying the layout to increase the area, multiple patterned layers can be used to color different nets and wherever required metal stitching can be used to short two colors (two nets). This is illustrated in Figure \ref{fig:both_stitch}. The metal stitched layout in figure \ref{fig:both_stitch}(b) permits a smaller spacing between the ZN, VDD, and VDD\_2 nets than would be possible if the ZN net were drawn with one color.  The downside of metal-stitching, however, is reduced predictabilty of wire parasitics and possibly increased chances of a manufacturing defect.

\section{Parasitic Extraction}
Another important component of the process design kit is the capability to correctly extract the interconnect and device parasitics. Parasitic extraction is an essential step in analyzing the performance of the design and parasitic capacitance and resistance of the layout are essentially defined by the following layer characteristics: 
\begin{enumerate}
\item Geometrical characteristics: Minimum-drawn widths, spacing and layer thickness, via enclosures, and trapezoidal shapes for layers.
\item Electrical properties: Resistivities (or sheet resistances), permittivities for various dielectric layers (dielectric constants), via and contact resistances.
\end{enumerate}

For FreePDK15, the layer definitions and the characteristics are defined in a technology file or .mipt file and the Mentor Grpahics' Calibre xCalibrate and Calibre xRC tools are utilized for parasitic extraction.

\subsection{Layer properties}

The values for widths and pitches for various metal layers were derived from the Interconnect tables from ITRS 2011 predictions for the 2016 node \cite{ITRS_interconnect} . However, ITRS-2011 predictions are more aggressive for metal1 and intermediate metal layer scaling than existing 15 nm processes \cite{Schuddinck:FinFET}. Therefore, the minimum width for the metal1 layer, which is often assumed to be roughly 1.5 times the minimum gate length, is assumed to be 28 nm, which is  twice that of the minimum gate length. Similarly, the dimensions of the intermediate metal layers are based on this assumption, while the semi-global and global layer dimensions are derived from the ITRS-2011 tables. The electrical and geometrical characteristics of the layer stack are listed in Table \ref{BEOL_table} and Table\ref{FEOL_table}. This stack was chosen to follow \cite{Schuddinck:FinFET} while filling in the gaps with materials that provide the approximate resistivity and dielectric constants predicted by ITRS. 

\begin{table}[htbp]
\centering
\caption{BEOL Layer properties}
\label{BEOL_table}
\includegraphics[width=0.9\linewidth, height = 0.8\linewidth]{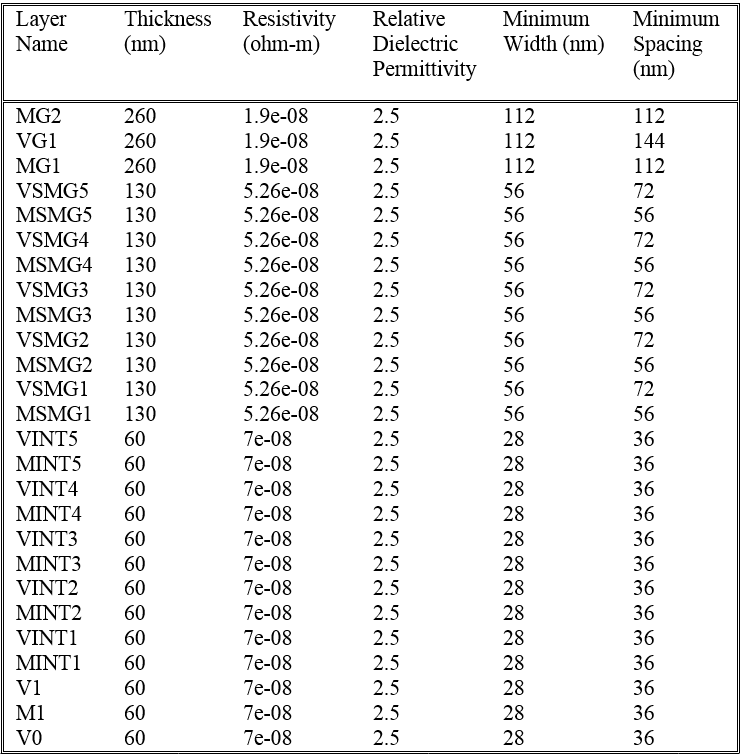}
\end{table}%

\begin{table}
\centering
\caption{FEOL and MOL Layer properties}
\label{FEOL_table}
\includegraphics[width=0.95\linewidth]{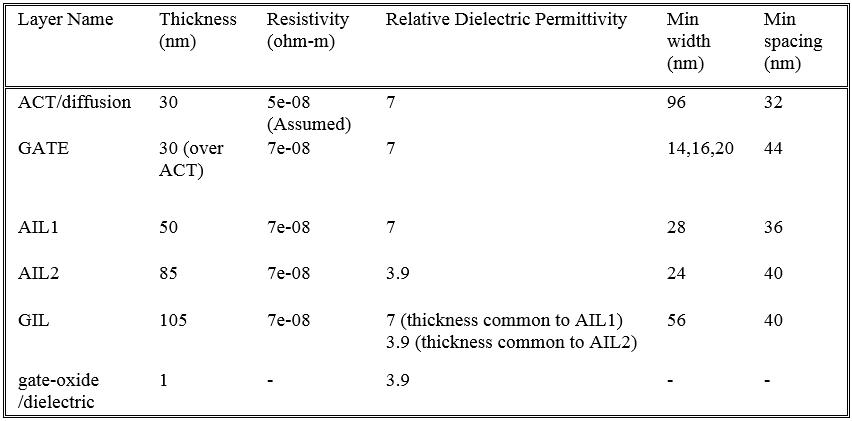}
\end{table}

\begin{figure*}
\centering
\begin{minipage}{.33\textwidth}
  \centering
  \includegraphics[width=1\linewidth]{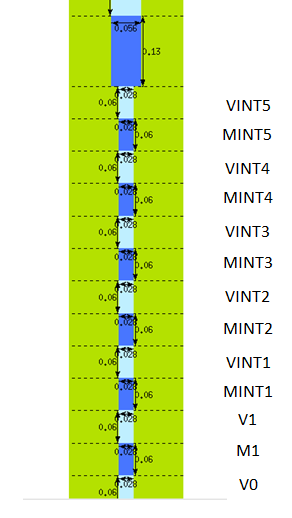}
  \subcaption{Intermediate metal layers and layer M1}
  \label{fig:beol1}
\end{minipage}%
\begin{minipage}{.33\textwidth}
  \centering
  \includegraphics[width=0.8\linewidth]{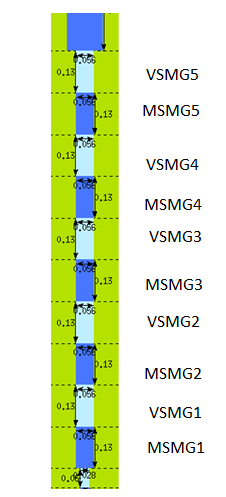}
  \subcaption{Semi-Global metal layers}
  \label{fig:beol2}
\end{minipage}
\begin{minipage}{.33\textwidth}
  \centering
  \includegraphics[width=.9\linewidth]{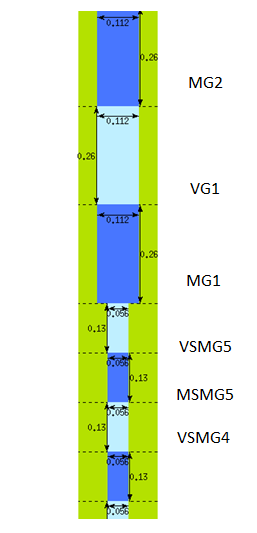}
  \subcaption{Global metal layers}
  \label{fig:beol3}
\end{minipage}
\caption{BEOL layer stack in Mentor Graphics' Stack Viewer}
\label{fig:both_beol}
\end{figure*}

\begin{figure*}
\centering
\begin{minipage}{.48\textwidth}
  \centering
  \includegraphics[width=0.9\linewidth]{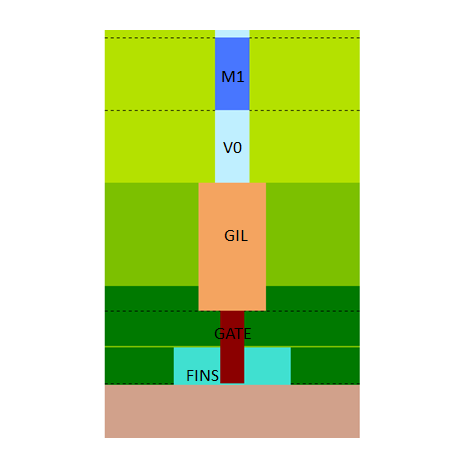}
  \subcaption{Stack with poly and GIL}
  \label{fig:feol1}
\end{minipage}%
\begin{minipage}{.5\textwidth}
  \centering
  \includegraphics[width=0.9\linewidth]{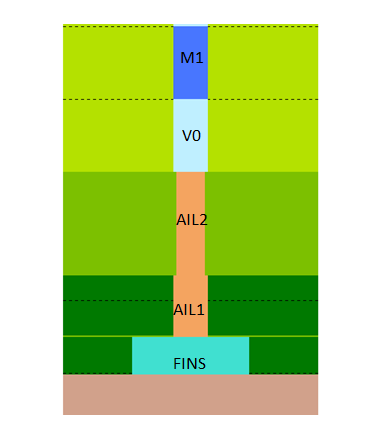}
  \subcaption{Stack with AIL1 and AIL2 }
  \label{fig:feol2}
\end{minipage}
\caption{FEOL and MOL layer stacks in Mentor Graphics' Stack Viewer}
\label{fig:both_mol}
\end{figure*}

Therefore, Tetraethyl Orthosilicate (TEOS) is selected as the dielectric surrounding the metal layers, while Silicon Nitride (SiN) is selected as the dielectric surrounding all the FEOL and MOL Layers, with the exception of AIL2 which uses silicon dioxide SiO\textsubscript{2}. Additionally, the top of GIL layer has same dielectric as AIL2 i.e. SiO\textsubscript{2} and bottom has the same dielectric as AIL1 i.e. SiN. 

The appropriate identification of the metal stack is validated by the output of the XCalibrate's stack-viewer tool shown in Figure \ref{fig:both_beol}, while that for the FEOL and MOL stack is shown in Figure \ref{fig:both_mol}.

\subsection{Capacitance extraction of FinFET}
The modeling of internal parasitics is highly complex due to the three-dimensional nature of the FinFET structure. However, the BSIM-CMG (for common multi-gate devices) \cite{BSIM} spice model developed by the BSIM group at UC Berkeley accounts for most of the internal device capacitances. However, the fin rises above the substrate resulting in additional capacitance with the external layers.

\begin{figure*}[htbp]
\centering
\begin{minipage}{.7\textwidth}
  \centering
  \includegraphics[width=0.975\linewidth]{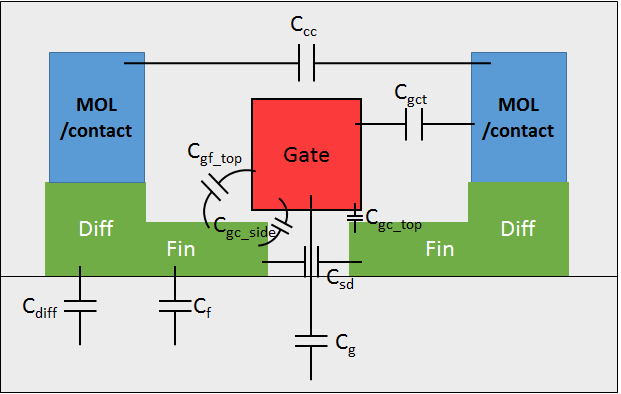}
  \subcaption{Cross-section view FinFET capacitance}
  \label{fig:cap_fin1}
\end{minipage}

\begin{minipage}{.7\textwidth}
  \centering
  \includegraphics[width=0.975\linewidth]{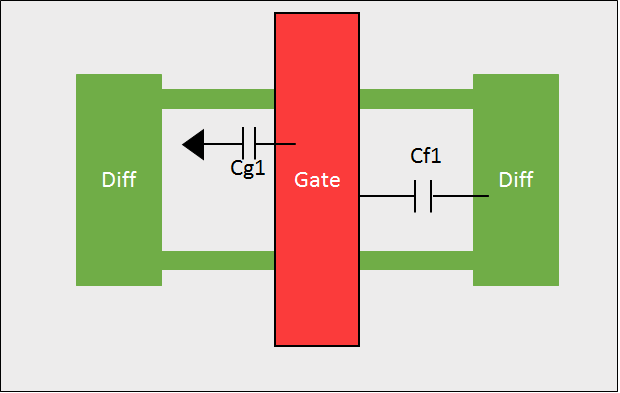}
  \subcaption{ Top view FinFET capaciitance}
  \label{fig:cap_fin2}
\end{minipage}
\caption{Capacitance components of FinFET divided between Spice model and extracted netlist \cite{LCollins}}
\label{fig_cap_fin}
\end{figure*}

\begin{table*}%
\centering
\caption{FinFET capacitances \cite{LCollins} \label{tab:one}}{%
\begin{tabular}{ l l }
\hline
\hline
Capacitances &  Domain\\ \hline
Contact to contact (Ccc)  &  Extraction \\ 
Gate to contact(C\textsubscript{gct}) & Extraction \\
Gate to top of fin(C\textsubscript{gf\_top}) & Extraction \\
Gate to substrate between fins(C\textsubscript{g1}) & Extraction \\
Gate to diffusion between fins(C\textsubscript{f1}) & Extraction \\
Source to drain(C\textsubscript{sd}) & Spice model \\
Gate to channel(C\textsubscript{g}) & Spice model\\
Fin to substrate(C\textsubscript{f}) & Spice model \\
Gate to fin inside channel(C\textsubscript{gc\_top} , C\textsubscript{gc\_side}) & Spice model \\ 
Diffusion to substrate(C\textsubscript{diff}) & Spice model \\
\end{tabular}}
\end{table*}%

Figure \ref{fig_cap_fin} illustrates various capacitances associated with a  FinFET device. As indicated in Table \ref{tab:one},  internal device capacitances like C\textsubscript{sd}, C\textsubscript{gc\_top} are accounted for in the BSIM-CMG \cite{BSIM} spice model. The other capacitances like contact-to-contact, gate to contact, and gate-to-substrate are accurately extracted through parasitic extraction process.

The comparison of this model with the FreePDK15 shows that the capacitances C\textsubscript{gf\_top}, C\textsubscript{f1}, C\textsubscript{cc} and C\textsubscript{gct} are extracted by the rules developed for FreePDK15. The capacitance C\textsubscript{gf\_top} is the fringing capacitance for gate over fins, C\textsubscript{f1} is the capacitance between GATE/GIL and AIL1. C\textsubscript{gct} and C\textsubscript{cc} are again GIL to AIL1 and GIL to M1 (in case of a direct connection) respectively. In FinFET layouts, fins are not represented as thin strips, however, the width of the active area is defined as the sum of fin width, and fin pitch times number of fins as shown in equation \ref{eq:fin}

\begin{equation}
W = W\textsubscript{fin} + (n\textsubscript{fin} - 1)*Pitch\textsubscript{fin}
\label{eq:fin}
\end{equation}

Thus, due to the way in which the width is defined, capacitance C\textsubscript{g1} is not currently modeled in FreePDK15. Additionally, these extracted capacitances have not been validated as that can only be achieved by comparing these results against results from a complex simulation using a 3D field solver. Thus, the current kit only represents an approximate value of the extracted FinFET capacitances.

\subsection{Validation of parasitic extraction}
The parasitic extraction process involves capacitive and resistive extraction of a given layout. The validation process includes design of simple layout and the comparison of their parasitics with first order models and approximations.
Capacitance validations include validation of parallel plate capacitance, fringing capacitance and coupling capacitance, while resistance validation includes comparison of sheet resistance. 

\subsubsection{Parallel plate approximation}
Capacitance between combination of metal layers of varying dimensions are compared against their parallel plate approximation model. It is found that for larger dimensions extracted capacitance for these metal layers matches the parallel plate capacitance. However, for smaller dimensions, (1um*1um) the difference between the extracted value and the estimated value was as high as 100\%. This difference is due to the fringing capacitance which is not included in the parallel plate model, and starts dominating at lower dimensions.

\subsubsection{Fringing capacitance modification}
In order to account for the Fringing capacitance, the parasitic capacitances are compared against the total capacitance values obtained from Sakurai's\cite{Sakurai} and Chang's\cite{Chang} approximation. It is found that the estimated parasitics have a significantly lower variation (\textless10\%) even for lower dimensions.

\subsubsection{Inclusion of coupling capacitance}
Coupling capacitance also significantly contributes to the overall parasitic capacitance of the layout, and thus in order to thoroughly validate it, the total  extracted capacitance must be compared with the model that accounts for the coupling capacitance. The process of validation thus involves modifying the dimensions and spacing between the metal layers and comparing that against Sakurai's approximation. It is found that the difference significantly improves (\textless2\%) and remains the same even when separation or lengths are modified.

\subsubsection{Resistance validation}
Simple sheet resistance formula was used to validate resistance extraction. The process involves varying the lengths and widths of different metal layer shapes and observing their effects on the extractes parasitic resistance. Sheet resistance values were calculated from the layer properties table and were used to validate the resistance values. However, in modern chips, the metal layer often is substituted for silicides or mixture of metals with varying quantities is used, which cannot give a simple value of resistivity. Moreover, with effects like skin effect at higher frequencies, resistance varies with distance from surface and hence parasitic resistance validation is kept to this simple sanity check.

\begin{figure}[!h]
\centering
\includegraphics[width = 0.48\textwidth]{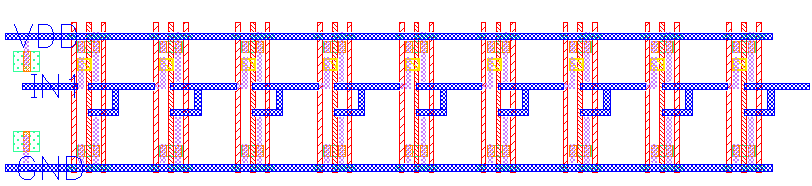}
\DeclareGraphicsExtensions{.pdf,.jpeg,.png,.jpg} 
\caption{ Chain of FO1 Inverters }
\label{fig_fo1}
\end{figure}  

\subsubsection{Inverter chain example}
To analyze and evaluate the validity of the kit, the propagation delay for a set of circuits were calculated and the results were compared with the estimated propagation delay from ITRS tables.

The delay analysis was performed for nine-stage FO1 and FO4 Inverters  and the average propogation delay for each single stage was calculated. Also, the technology models used for the HSPICE simulations were PTM’s 14 nm High Performance nfet and pfet models \cite{PTM}, which are based on the BSIM models for common multi-gate devices \cite{BSIM}. 

In order to study the impact of additional parasitics on these models the following the propagation delay was computed for each of the following cases
\begin{enumerate}
\item Basic circuit based on only the spice models
\item Circuit with source and drain dimensions defined. This enables inclusion of parasitic capacitance in the spice model.
\item Circuit with extracted parasitic netlist for the corresponding circuit layout.
\end{enumerate}

From the delay analysis performed for these cases, it is found that the addition of spice model parasitics as well as inclusion of the parasitic extraction results increases the propagation delay of the circuit. This is in agreement with the estimated behavior of the circuit.
Furthermore, another objective of the delay analysis is to evaluate whether the results obtained from the simulations have the same order of magnitude and are within the neighborhood of the propagation delay predicted by ITRS. The simulation result for FO1 Inverter is 1.81 ps, while that predicted by ITRS is 3 ps. Similarly, the result obtained for FO4 Inverter is 4.39 ps, while that predicted by ITRS is 7.18 ps \cite{ITRS:Online}. This indicates that results are close to the predicted values.
\begin{table*}[htbp]%
\centering
\caption{Propogation delay for Inverter chains \label{tab:inv}}{%
\begin{tabular}{ l l l l}
\hline
\hline
Circuit &  tp - PTM {$^a$}(ps) & tp - S/D specified{$^b$} (ps) & tp -  extracted{$^c$} (ps)\\ \hline
FO1 Inverter  &  1.28 & 1.75 & 1.80\\
FO4 Inverter  & 3.94 & 4.30 & 4.39 \\
\end{tabular}}\\
\footnote{$^a$}{Single stage propagation delay for circuit based on spice models.}\\
\footnote{$^b$}{Single stage propagation delay for circuit with source and drain dimensions specified}\\
\footnote{$^c$}{Single stage propagation delay for circuit with layout extracted netlist}\\
\end{table*}%

\section{Conclusions}
The introduction of integrated circuit design using FinFET devices in university education is currently constrained due to high licensing cost of the commercial design flows. FreePDK15 attempts to remove this constraints by providing an open source predictive process design flow platform wherein circuits for 15 nm FinFET devices can be designed and verified. 
In this paper, a PDK is described which consists of a layer stack based on existing FinFET designs and ITRS predictions. The design rules encompassing special rules for double patterning lithography, gate cut layers and MOL layers are implemented.

Since the geometrical characteristics of a FinFET layout differ from that of a planar MOSFET, the modifications required for correctly extracting the source and drain dimensions (accounting for the number of fins) and FinFET device recognition are executed. Additional rules requiring layout extraction of interconnects because of double patterning and metal stitching are introduced and validated. 

A study of the parasitic capacitance components of the FinFET device indicates additional capacitances due to the folding of gate over channel. However, due to the manner in which the current layouts are drawn all the parasitic components have not been accounted for and would require furtherl complex modeling of the 3D gate structure. However, the interconnect capacitances and resistances are validated against standard models and are found to be closer to the estimated values. The complete design flow is proven by running simulations on extracted netlists of FO1 and FO4 Inverters and the propagation delay results of these simulations were found within the vicinity of the results predicted by ITRS-2011 tables for 2016 node.

\section*{Acknowledgment}
The authors would like to thank Paul Franzon at NC State University. The authors would like to thank Mentor Graphics, since this project would not have been possible without their generous  gift  of  supporting  funds  and  Calibre  licenses. The authors  would  also  like  to  thank  Tarek  Ramadan,  Ahmed Hammed  Fathy,  Omar  El-Sewefy,  Ahmed  El-Kordy,  Hend Wagieh and the team at Mentor Graphics for development of the first  set  of design rules and their constant  support.In addition, the authors would like to thank and acknowledge Alexandre Toniolo at Nangate for clarifying the vision of MOL layers. We would also like to thank Cadence designsystems for use of the virtuoso software and Synopsys Inc.for use of Pycell studio. The authors would also like to thanks Vikas Sharma for P-Cells, Vidyanandgouda Patil for design rule fixes and Namrata Sampat for help cleaning up the distribution.

\ifCLASSOPTIONcaptionsoff
  \newpage
\fi



%
\bibliographystyle{IEEEtran}  
\bibliography{IEEEabrv,bibi} 


%

%







\end{document}